\documentclass[iop]{emulateapj}

\usepackage{epsfig}
\usepackage{natbib}
\def\simgt{\lower.5ex\hbox{$\; \buildrel > \over \sim \;$}}
\def\simlt{\lower.5ex\hbox{$\; \buildrel < \over \sim \;$}}

\def\amin{\ifmmode^{\prime}\else$^{\prime}$\fi}
\def\asec{\ifmmode^{\prime\prime}\else$^{\prime\prime}$\fi}

\def\simgt{\lower.5ex\hbox{$\; \buildrel > \over \sim \;$}}
\def\simlt{\lower.5ex\hbox{$\; \buildrel < \over \sim \;$}}

\newcommand{\loe}{\stackrel{<}{\sim}}
\newcommand{\goe}{\stackrel{>}{\sim}}

\newcommand\chandra{{\it Chandra}}

\newcommand\swift{{\it Swift\/}}
\newcommand\nustar{\hbox{\it NuSTAR\/}}

\def\pdot{$\dot P$}
\def\edot{$\dot E$}

\def\sga{Sgr A*}

\def\mgt{SGR~J1745$-$29}

\slugcomment{Accepted to The Astrophysical Journal Letters}

\shorttitle{Discovery of magnetar in Sgr A* region}
\shortauthors{}

\begin{document}

\title{NuSTAR discovery of a 3.76-second transient magnetar near Sagittarius A*}

\author{ Kaya Mori\altaffilmark{1}, Eric V. Gotthelf\altaffilmark{1}, Shuo Zhang\altaffilmark{1},
  Hongjun An\altaffilmark{2}, Frederick K. Baganoff\altaffilmark{3}, Nicolas M. Barriere\altaffilmark{4}, Andrei M. Beloborodov\altaffilmark{1},
  Steven E. Boggs\altaffilmark{4}, Finn E. Christensen\altaffilmark{5}, William W. Craig\altaffilmark{4}, Francois Dufour\altaffilmark{2},
  Brian W. Grefenstette\altaffilmark{6}, Charles J. Hailey\altaffilmark{1}, Fiona A. Harrison\altaffilmark{6}, Jaesub Hong\altaffilmark{7},
  Victoria M. Kaspi\altaffilmark{2}, Jamie A. Kennea\altaffilmark{8}, Kristin K. Madsen\altaffilmark{6}, Craig B. Markwardt\altaffilmark{9},
  Melania Nynka\altaffilmark{1}, Daniel Stern\altaffilmark{10}, John A. Tomsick\altaffilmark{4}, William W. Zhang\altaffilmark{9}}

\altaffiltext{1}{Columbia Astrophysics Laboratory, Columbia University, New York, NY 10027, USA; kaya@astro.columbia.edu}
\altaffiltext{2}{Department of Physics, McGill University, Montreal, QC H3A2T8, Canada}
\altaffiltext{3}{Kavli Institute for Astrophysics and Space Research, Massachusets Institute of Technology, Cambridge, MA 02139, USA}
\altaffiltext{4}{Space Sciences Laboratory, University of California, Berkeley, CA 94720, USA}
\altaffiltext{5}{DTU Space - National Space Institute, Technical University of Denmark, Elektrovej 327, 2800 Lyngby, Denmark}
\altaffiltext{6}{Cahill Center for Astronomy and Astrophysics, California Institute of Technology, Pasadena, CA 91125, USA}
\altaffiltext{7}{Harvard-Smithsonian Center for Astrophysics, Cambridge, MA 02138, USA}
\altaffiltext{8}{Department of Astronomy and Astrophysics, The Pennsylvania State University, University Park, PA 16802, USA}
\altaffiltext{9}{NASA Goddard Space Flight Center, Greenbelt, MD 20771, USA}
\altaffiltext{10}{Jet Propulsion Laboratory, California Institute of Technology, Pasadena, CA 91109, USA}

\begin{abstract}

We report the discovery of 3.76-s pulsations from a new burst source
near Sgr A* observed by the
\nustar\ Observatory.  The strong signal from \mgt\ presents a
complex pulse profile modulated with pulsed fraction $27\pm3$\% in the
$3-10$~keV band.  Two observations spaced 9~days apart yield a spin-down rate of \pdot
$=(6.5\pm1.4)\times10^{-12}$. This implies a magnetic field
$B=1.6\times10^{14}$ G, spin-down power \edot $=
5\times10^{33}$~erg~s$^{-1}$, and characteristic age $P/2$\pdot $=
9\times10^3$ yr, for the rotating dipole model. However, the current \pdot\ may be erratic, especially during outburst.
The flux and modulation remained steady during the observations and the
$3-79$~keV spectrum is well fitted by a combined blackbody plus
power-law model with temperature ${\it k}T_{\rm BB} = 0.96\pm0.02$ keV
and photon index $\Gamma = 1.5\pm0.4$. 
The neutral hydrogen column density ($N_H\sim1.4\times10^{23}$ cm$^{-2}$) measured
by \nustar\ and \swift\ suggests that \mgt\ is located at or near the Galactic Center.    
The lack of an X-ray counterpart in the published
\chandra\ survey catalog sets a quiescent $2-8$~keV luminosity limit of $L_{x}                 
\simlt 10^{32}$~erg~s$^{-1}$. The
bursting, timing, and spectral properties indicate a
transient magnetar undergoing an outburst with $2-79$~keV luminosity up to 
$3.5\times10^{35}$~erg~s$^{-1}$ for a distance of 8 kpc. \mgt\ joins a growing subclass of transient magnetars, indicating that many magnetars in quiescence remain undetected in the X-ray band or have been detected as high-B radio pulsars.   
The peculiar location of \mgt\ has important implications for the formation and dynamics of neutron stars in the Galactic Center region.  

%Based on the observations of other transient                                        
%magnetars (1E1547.0-5408, J1622-4950 and XTE J1810-197), \mgt\ may                  
%remain bright in the X-ray band with a decay time of over an year.                  

\end{abstract}

\section{Introduction}

The small class of young neutron stars that exhibit sudden bright
X-ray and soft gamma-ray bursts, large X-ray flares and often
strong, broad X-ray pulsations are believed to be ``magnetars'':
neutron stars whose radiation is powered by the decay of intense magnetic fields
\citep{td95,td96,tlk02,bel09}.  These high fields are inferred
independently of energetics from their measured
spin-periods $P$ and spin-down rates $\dot{P}$, under the standard
assumption of magnetic dipole braking in a vacuum.  With the
currently known objects numbering only two dozen\footnote{See the
  complete online magnetar catalog at
  http://www.physics.mcgill.ca/\~{}pulsar/magnetar/main.html}, and with this group exhibiting a wide variety of
interesting, often dramatic phenomena \citep{woods2006,mereghetti2013}, the
physics of magnetars is still
poorly understood.  Each new magnetar potentially
provides another important piece of the neutron star puzzle.

%[THIS PARAGRAPH MATERIAL MAINLY BELONGS IN THE OBSERVATION SECTION...]              
%\nustar\ is the most recently launched X-ray mission and the first                  
%focusing hard X-ray telescope \citep{hcc+13}.  Launched in June 2012,               
%\nustar\ is sensitive in the 3-79 keV band, with angular resolution                 
%18$''$ (FWHM).  Moreover \nustar\ has 2-$\mu$s time resolution, very                
%useful for observations of pulsed X-ray sources.  \nustar's nominal                 
%2-yr baseline mission includes a significant component of magnetar                  
%observations \citep{kab+13} as well as a Galactic Center survey and                 
%monitoring program \citep{???}.                                                     

On 2013 April 24 UT, \swift\ monitoring of the Sgr A* region
revealed a large X-ray flare \citep{atel5006}.  On
April 26, the \swift/BAT instrument detected a short ($\sim$32~ms)
X-ray burst consistent with that from a magnetar. This event triggered
an immediate follow-up using \swift's X-Ray Telescope (XRT) that
localized a new point source at a position consistent with that of Sgr
A* \citep{atel5009}.
%to within its 90\% confidence error radius of 2.8$''$                               
\nustar\ initiated a
target-of-opportunity observation on April 26 that revealed 3.76-s
pulsations for the new source \citep{atel5020}.  The combination of a magnetar-like burst,
periodicity and spectrum led to the identification of the transient
as a likely new magnetar in outburst.  A \chandra/HRC-S observation made on
April 29 confirmed the pulsations and localized the pulsar to be
$\sim3''$ from the position of Sgr A* \citep{atel5032}.  Radio observations also 
detected pulsations at the X-ray period \citep{atel5040}, implying a dispersion measure range consistent with a position at the Galactic Center \citep{atel5043, atel5058, atel5064}, while the $N_H$ measured in the X-ray band is consistent with a position at the Galactic Center or slightly beyond.

% although an initial claim of a
%surprisingly low dispersion measure toward the source \citep{atel5035} has very recently
%been superseded by claimed detections at much a higher value, more consistent
%with the large $N_H$ \citep{atel5040,atel5043}.

%[THE RADIO RESULT SEEMS A BIT IFFY FOR NOW, MAYBE SHOULD QUALIFY UNITIL THIS SORTED OUT? THIS COULD GO IN THE DISCUSSION INSTEAD?].                                     
%Three follow-up radio observations at the                                           
%coordinates of the \chandra\ source strongly suggest a radio detection              
%of the 3.76-s pulsation. However the possibility of interference                    
%cannot yet be ruled out.                                                            
%While one radio observation at 3094 MHz                                             
%measured the dispersion measure $\loe 50$pc cm$^{-3}$, more recent                  
%radio observations at 4.85 and 8.35 GHz measured significantly larger               
%dispersion measure ($>10^3$pc cm$^{-3}$) indicating that \mgt\ is                   
%located at the Galactic Center region [ref to radio ATELs].                         
%If these radio detection are confirmed then an accreting                            
%neutron star origin for the pulsations can be rules out.                            

In this Letter, we detail the \nustar\ discovery of this new magnetar
and a \swift\ observation obtained 7.3~days later that provides a
confirming spin-down measurement. In \S2 we describe
the \nustar\ observations, in \S3 we present the pulsar
discovery and in \S4 the spectral analysis. Finally, in
\S5 we discuss the implications of a magnetar close
to the Galactic Center. A companion paper to this one focuses on the detection of the discovery of the SGR burst and monitoring its flux evolution \citep{kennea2013}.

%%%%%%%%%%%%%%%%%%%%%%%%%%%%%%%%%%%%%%%%%%%%%%%%%%%%%%%%%%%%
\section{NuSTAR Observations}

Following the \swift/BAT report of flaring activity from the Galactic
Center \citep{atel5006}, \nustar\ initiated a ToO observation on 2013
April 26, 1:17:31 UT.  This observation had 94.5 ks of exposure.
A second observation was carried out on May 4, at UT 17:49:21 for 42.0 ks.
In both observations, the Galactic Center region was imaged with the two
co-aligned X-ray telescopes on-board \nustar\ with the \sga\ source placed at the optical
axis. 
These mirror/detector units provide $58^{\prime\prime}$ (Half-power diameter) and $18^{\prime\prime}$ (FWHM)  
imaging resolution over the $3-79$~keV X-ray band,
with a characteristic spectral resolution of 400~eV (FWHM) at 10~keV \citep{harrison2013}. 

The nominal reconstructed \nustar\ coordinates are accurate to
$7.5^{\prime\prime}$ (90\% confidence level). Time tagged-photon arrival times are
accurate to $<2 \; \mu$s; the precise time resolution depends on the count rate incident
on the detector.  For the observations reported on here, deadtime is unimportant.
The absolute timing
accuracy of the \nustar\ time stamps is limited to \hbox{$\sim2$~ms} rms after
calibrating the thermal drift of the on-board clock.

Data were reduced and analyzed using the \nustar\ {\it Data Analysis                 
  Software (NuSTARDAS)} v10.1 in conjunction with FTOOLS 6.13. The
data were filtered for intervals of high background. 
%A total of 58.5-ks good on-source
%exposure time was collected.
%\nustar\ detected the bursting source                                               
% [WE DON'T KNOW THIS FOR SURE, WE DID NOT SEE BURSTS? SEE NEXT PARAGRAPH]           
%at RA=17:45:40.65, DEC=-29:00:27.45, consistent with the                            
%within the \nustar\ positioning uncertainty.                                        
%reported \chandra\ position reported by Rea et al.\citep{atel5032}                   
Photon arrival times were corrected to the Solar System barycenter
using the \chandra\ coordinates reported by \citet{atel5032}. 
Examination of the count rate in the \nustar\ image at the
\chandra\ reported coordinates of the burst shows clear evidence for
enhanced X-ray emission in the region. Data extracted from a
$1^{\prime}$ radius aperture in the $3-10$~keV and $10-79$~keV bands
yield count rates of $0.730\pm0.002$ cts~s$^{-1}$ and $0.099\pm0.002$ cts~s$^{-1}$,
respectively, $2.6$ and $1.2$ times higher than those of the pre-flare
observations. The pre-flare background rates were established using
three recent \nustar\ Galactic survey observations acquired in 2012
July, August and October.

%%%%%%%%%%%%%%%%%%%%%%%%%%%%%%%%%%%%%%%%%%%%%%%%%%%%%%%%%%%%%%%%%%%%%%%%%%%%

\section{Timing analysis}

To search for pulsations, we used an initial 7~ks of data acquired in the first ToO pointing. A
total of 16,500 photon arrival times were extracted in the full energy bandpass using a $30^{\prime\prime}$ radius aperture
centered on the burst location. The arrival times were binned at 2~ms
and searched for coherent pulsations up to the Nyquist frequency using
a $2^{22}$ FFT. We found a complex signal with three highly
significant Fourier components at 1.25 s, 3.76 s, 1.88 s, and 0.940 s,
ordered by decreasing strength.

\begin{figure*}[t]
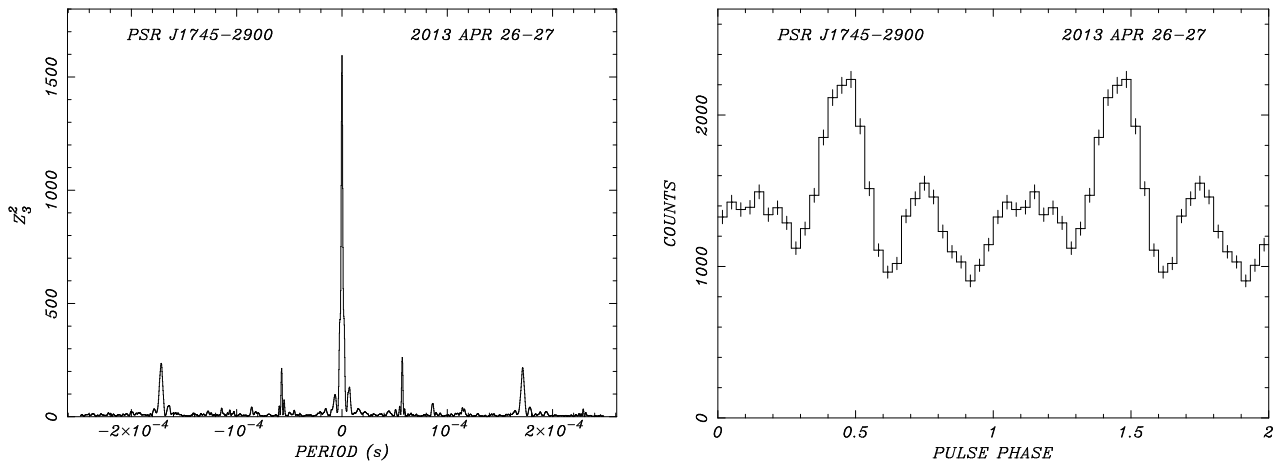

\centerline{
\hfill
\psfig{figure=sgra_pulsar_all_pgram.ps,height=0.45\linewidth, angle=270}
\hfill
\psfig{figure=sgra_pulsar_all_fold_bsub.ps,height=0.45\linewidth, angle=270}
\hfill }
\caption{Discovery observation of X-ray pulsations from \mgt\ using data collected
from a 94.5~ks \nustar\ observation of Sgr A*. Left -- The $Z^2_3$-statistic periodogram
shows a highly significant detection around $P = 3.7635$~s. Side-lobe aliases from the satellite 
orbit are evident. Right --- The light curve folded on the peak period of the periodogram.
Two cycles are shown for clarity. The phase offset is arbitrary.}
\label{fig:timing}
\end{figure*}

We then carried out a refined $Z^2_{\rm n}$ analysis using the entire
94.5~ks, restricting the energy band to $3-10$~keV above which
the source photons are dominated by the quiescent background. 
This allowed us to identify the fundamental at 3.76~s
with power at odd harmonics.  This signal corresponds to a pulse
profile with three resolved peaks each 0.6 s wide, dominated
by a single strong peak (see Figure~\ref{fig:timing}). A $Z^2_3$
analysis yields a period $P=3.76354455(71)$~s at Epoch MJD(TDB)
56409.2657 where the 1$\sigma$ error on the least significant digits
is given in parentheses.  The uncertainty is estimated from a Monte
Carlo simulation of the lightcurve using the method described by
\citet{gotthelf1999}.

The intrinsic pulsed fraction in the $3-10$~keV band is $f_p = 27\pm3\%$
after allowing for the background level in the source aperture,
estimated using data from the earlier observations. Here, we define the
pulse fraction as the ratio of the pulsed emission to the source
(background subtracted) flux. To determine the unpulsed level, because
of the high counting statistics, we take the lowest bin in the well
resolved 30-bin folded lightcurve shown in Figure~\ref{fig:timing}.
%The possibility of energy dependence of the pulse shape and pulsed                                                                  
%fraction was examined in narrow $1$~keV energy bands.                                                                               
The pulse shape shows little change with energy below 7~keV,
within statistics. Above $7$~keV, the main pulse sharpens
and the smaller sub-pulses becomes lost in the increasing background counts.
%The pulse fraction slowly increases with energy, with a                                                                             
%dip around [$6-7$~kev and $9-10$~keV??].                                                                                            
%(need more info!!)                                                                                                                  

Following the \nustar\ detection of \mgt, we requested a
\swift\ ToO observation to monitor its temporal and spectral
evolution. A total of 15.5~ks of X-ray data were collected on 2013 May
3 starting at UT 10:02:43.72 using the \swift/XRT (Burrows et
al. 2005) in windowed timing (WT) mode. In this mode, the XRT is
sensitive to photons in the $0.2-10$~keV band with 1.77-ms time
resolution. The XRT quicklook data were processed with {\tt
  xrtpipeline} (v. 0.12.6) and photon arrival times were corrected to
the barycenter using the \chandra\ coordinates. From a total of 4,900
counts collected from the source in a $0.4^{\prime}$ radius aperture
in the $0.3-10$ keV XRT bandpass we detect the pulsar signal with high
significance and measure a period of 3.7635603(68)~s at epoch MJD(TDB) 
56415.4186. A preliminary period derivative obtained by combining
this value with the above \nustar\ measurement was reported in
\cite{atel5046}.

%  P = 3.76354428(13)  Pdot = (6.1+/-1.9)E-12 tau = 9.7 kyr B = 1.5e14 G Edot = 4.6e33 erg/s                                         

Adding the two \nustar\ pointings and the \swift\ data together, we
searched over ($f,\dot f$)-space around the initial values reported in
\cite{atel5046} using the $Z^2_3$ statistic to derive a revised
spin-down rate of $\dot P = (6.5\pm1.4) \times 10^{-12}$, which, taken
at face value, implies a magnetic field $B = 1.6\times10^{14}$~G,
spin-down power $\dot E = 5\times 10^{33}$~erg~s$^{-1}$, and characteristic
age $P/2\dot P = 9$~kyr, assuming a vacuum dipole. A similar $\dot P$ value ($6.5\times10^{-12}$) was measured by a follow-up radio observation of \mgt\ \citep{atel5058}. $\dot{E}$ is
smaller than the concurrent X-ray luminosity by orders of magnitude,
which rules out rotation power as the source of the X-ray emission. We note that the spin-down
rate of magnetars can be highly variable, especially following
outburst \citep{kaspi2003, dib2008}.

%moreover the current value of $\dot P$ may be significantly larger                                                                  
%than the long-term spin-down rate due to the possibility of recovery                                                                
%from a glitch, as is commonly observed in magnetars in outburst                                                                     

We also searched the \nustar\ $3-79$~keV lightcurves over a range of
timescales for magnetar-like bursts similar to the 32\,ms burst
reported in \citet{atel5009}. 
A burst with the reported properties would have easily been detected in our
data. However, we did not detect any statistically significant bursts.

\section{Spectral analysis}

We analyzed the full spectral data from the first \nustar\ observation, which consists of two consecutive data sets (OBSID 30001002006 and 80002013002) for a total of $94.5$~ks. The extraction region of 1$^{\prime}$ in radius encompasses strong diffuse emission and numerous unresolved sources within the Sgr A complex, so we extracted the background from a past observation in which \mgt\ was not detected and \sga\ did not exhibit any detectable flare (OBSID 30001002003 from 2012 August 4 at UT 07:56 to August 6 at UT 01:06).

Joint fitting with \swift\ was conducted to better constrain the column density. Five \swift/XRT observations which covered the first \nustar\ observation window were used (\swift\ Seq\# 00554491001, 0009173620, 0009173621, 00554491991 and 00035650242), yielding 26-ks exposure time in total. The data were reduced with {\tt xrtpipeline}. A 22$^{\prime\prime}$ radius aperture was used to extract source photons, and the background contribution was estimated by extracting photons from a concentric annulus of inner radius $70^{\prime\prime}$ and outer radius $160^{\prime\prime}$. 

Joint spectral analysis was done in the $1.7-8.0$ keV energy band for Swift data and $3-79$ keV for \nustar\ data using XSPEC \citep{arnaud.96}, setting the atomic cross sections to \cite{verner.96} and the abundances to \cite{wilms.00}. Table \ref{tab:specfit} shows the results. The low energy spectrum is well fit by an absorbed blackbody (BB), but the high energy tail clearly requires the addition of a power-law (PL) component. 
The model $\texttt{TBabs} \times (\texttt{bbody} + \texttt{pegpwrlw})$ yields a reduced $\chi^2$ of 1.01. 
The following fluxes were extracted using the convolution model $\texttt{cflux}$ for the best-fit BB + PL model over the joint energy band; 
the absorbed flux is $(2.67 \pm 0.02) \times 10^{-11}$~ erg~cm$^{-2}$~s$^{-1}$, and the unabsorbed flux is $(4.55 \pm 0.04) \times 10^{-11}$~ erg~cm$^{-2}$~s$^{-1}$. 
Placing the source at the Galactic Center (distance of 8 kpc, \citet{reid1993}), the inferred $2-79$ keV luminosity is $3.5  \times 10^{35}$~erg~s$^{-1}$.

\begin{deluxetable}{lcccc}
%\tabletypesize{\small}
\tablecaption{Spectral modeling of the {\it Swift} and {\it NuSTAR} data.} 
\tablecolumns{2}
%\tablewidth{0pc}
\tablehead{ \colhead{Model}   &  \colhead{BB}  &  \colhead{BB+PL} }
\startdata
$N_{\rm H}$ (10$^{22}$ cm$^{-2}$) & $12.98^{+0.54}_{-0.52}$ & $14.20^{+0.71}_{-0.65}$ \\
$kT$ (keV)   &  $1.000 \pm 0.010$   & $0.956^{+0.015}_{-0.017}$  \\
BB flux (erg~cm$^{-2}$~s$^{-1}$) & $(4.39 \pm 0.04) \times 10^{-11}$  & $(4.73 \pm 0.04) \times 10^{-11}$ \\
BB luminosity (erg~s$^{-1}$) & $\cdots$ & $(3.62 \pm 0.03) \times 10^{35}$\\ 
BB radius (km) & $\cdots$ & $1.7 \pm 0.1 $\\ 
$\Gamma$  & $\cdots$ &  $1.47^{+0.46}_{-0.37}$ \\
PL flux (erg/cm$^2$/s) & $\cdots$ & $(6.22 \pm 0.57) \times 10^{-12}$ \\
$\chi^2_{\rm r}$ (dof)   & 1.44 (466) & 1.01 (464) \\
\enddata
\tablecomments{$N_{\rm H}$ is the column density, $kT$ is the temperature of the blackbody, $\Gamma$ is the photon index of the power-law. The $2-79$ keV fluxes are given for the individual components. The goodness of fit is evaluated by the reduced $\chi^2$, and the degrees of freedom is given between brackets. The errors are 90\% confidence ($\Delta\chi^{2} = 2.7$). The blackbody radius is assuming a distance of $8$~kpc.}
\label{tab:specfit}
\end{deluxetable}

We also investigated any phase dependence of the \nustar\ spectrum by segmenting the data into 6 non-overlapping intervals, consisting of three peak and three off-peak regions.  
The background spectrum was identical to what was used in the phase-averaged spectral analysis detailed above, and was corrected to account for the phase cuts. A consequence of such fine division is poor photon statistics above $\sim10$ keV.  Accordingly, the phase-resolved spectra were only able to constrain an absorbed black body model with a fixed column density.  
The phase-resolved spectra were fit with an absorbed blackbody model holding $N_H$ fixed to 13$\times10^{22}$ cm$^{-2}$.  We found a $4\%\pm2\%$ variation in $kT$ while the blackbody flux normalization varied by $\sim30$\%.
%We conclude that the flux modulation \ref{fig:timing} is primarily due to the flux variation with little temperature variation. 

\begin{figure}[t]

\psfig{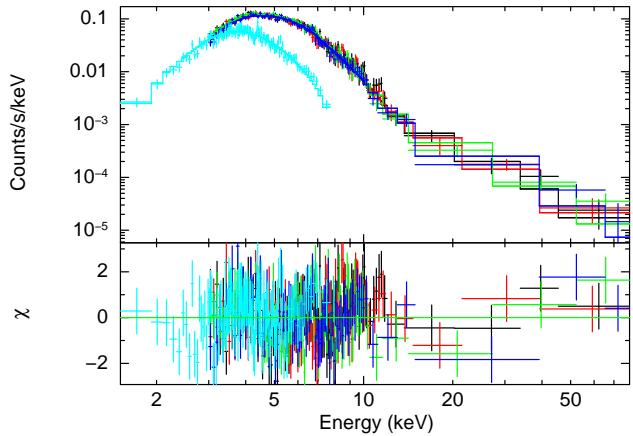}

\caption{\nustar\ (black and red for telecope module A and B, respectively, from OBSID 30001002006, and green and blue for telescope module A and B, respectively, from OBSID 80002013002), and \swift\ (cyan) spectra jointly fitted to an absorbed blackbody plus power-law model. The crosses show the data points with 1-$\sigma$ error bars, and the solid lines show the best fit model. The lower panel shows the deviation from the model in units of standard deviation.}
\label{fig:image3}
\end{figure}

%%%%%%%%%%%%%%%%%%%%%%%%%%%%%%%%%%%%%%%%%%%%%%%%%%%%%%%%%%%%%%%%%%%%%%%%%%%                                           

%\section{Burst search}

%For $N_H = 1\times 10^{23}$\,cm$^{-2}$, 
%{\bf PROBLEM AS ABOVE YOU SHOW NH>1e23 -- VK}
%such a burst would 
%produce $\sim$20 {\em NuSTAR} counts, i.e. would have been
%easily detected.   
%Other spectral 
%models, such as thermal bremsstrahlung \citep{woods2006}  
%give significantly higher count estimates.  
%For any typical 
%model, {\em NuSTAR} would have detected a burst as bright as
%the one reported.
% ZAPPED ABOVE 2 SENTENCES TO SAVE SPACE -- VK

%{\bf YOU COULD SAVE SPACE ABOVE BY NOT HAVING ALL THE SUBSECTIONING; IT'S NOT
%SO CRUCIAL IN A SHORT LETTER -- VK}

%%%%%%%%%%%%%%%%%%%%%%%%%%%%%%%%%%%%%%%%%%%%%%%%%%%%%%%%%%%%%%%%%%%%%%%%%%%%%%%%%%

\section{Discussion}

%\subsection{Is \mgt\ a transient magnetar?}  

The {\it Swift}-observed magnetar-like burst, the detected spin period (3.76 s) and the preliminary first derivative as measured by \nustar\ and {\it Swift} provide firm evidence that \mgt\ is a new magnetar in outburst. 
%with an inferred dipole B-field strength of $B\sim3\times10^{14}$ G, spin-down power of $\sim2\times10^{34}$ erg/s and spin-down age of $\sim2\times10^3$ years. 
%{\bf NO!  THE B COULD BE SIGNIFICANTLY BIASED.  YOU CANNOT SAY THE ABOVE WITH ANY CERTAINTY -- VK}
\mgt\ has shown no significant flux variation over $\sim10$ days since the burst was detected on 2013 April 24 \citep{kennea2013}.
%\nustar\ has thus far seen no time variability in the X-ray spectra and flux.

In archival data, there is no X-ray counterpart at the \chandra\ position of \mgt. This sets an upper limit on the quiescent $2-10$ keV luminosity of $\loe10^{32}$~erg~s$^{-1}$ \citep{muno2009} while the bursting $2-10$ keV luminosity of \mgt\ reached $\sim3\times10^{35}$~erg~s$^{-1}$.  Comparable dynamic ranges have been seen in other magnetars,
including 1E~1547$-$5408 \citep{sak2011} and Swift J1822.3$-$1606 (e.g. \citet{scholz2012}); such sources
have been dubbed ``transient'' (e.g. \citet{halpern2008}) to distinguish them from the ``classical'' magnetars with high quiescent luminosities (see \citet{kaspi2010} and references therein). With its detection at radio wavelengths \citep{atel5040,atel5043}, \mgt\ is similar to 
the transient magnetars 1E 1547$-$5408 \citep{camilo2007}, J1622$-$4950 \citep{levin2010} and XTE J1810$-$197 \citep{camilo2006}, the only three radio-detected magnetars. 
That all radio-detected magnetars are transients in spite of deep radio searches of classical magnetars \citep{burgay2006,crawford2007,lazarus2012} suggests the radio emission may be associated only with the transients.
The lack of a quiescent \chandra\ counterpart indicates that \mgt\ has a quiescent temperature of $kT \loe$ 0.3 keV, assuming that the source is at the Galactic Center (8 kpc). 
This is comparable to the quiescent temperature of the transient magnetar XTE J1810$-$197 \citep{gotthelf2004}, and, interestingly, to the blackbody temperatures of a growing
number of high-magnetic-field radio pulsars \citep{olausen2013}.  This fact, and the detection of a magnetar-like outburst from one high-B pulsar \citep{gavriil2009}, suggests that all high-B radio pulsars may be magnetars in quiescence, consistent with models of magnetothermal evolution (e.g. \citet{ponsp2011}; \citet{pernap2011}).
%and the thermal emission comes from the whole 10 km-in-radius surface. 
%We did not detect any additional thermal emission, while the 2--10 keV X-ray spectrum is dominated by a $kT \sim 1$ keV blackbody.

%The post-burst emission can be powered by two sources --- (1) heat generated in the deformed crust of the neutron star and (2) magnetospheric activity caused by the deformation of magnetic %field lines. The nonthermal spectrum of SGR J1745-2900 points to the magnetospheric origin. 

The hard X-ray power-law component with slope $\Gamma\sim 1$ is similar to what has been observed for other magnetars at high energies (e.g. \citet{denHartog2008, enoto2010}).  It is suggested to be generated by an electron-positron flow in a closed twisted bundle of magnetic field lines \citep{belobor2013}.  In this model, the flow is sustained through $e^\pm$ discharge at voltage $\Phi=10^9-10^{10}$~V, and energy is released with rate $L=I\Phi$ where $I$ is the electric current circulating in the twisted bundle.
If the bundle is near the magnetic dipole axis and emerges from a polar cap of area $A$,
its magnetic flux is $AB$ (where $B$ is the surface magnetic field and $A = 10^{11.5}A_{11.5}$~cm$^2$), and it generates
luminosity $L\sim 10^{35}\mu_{32}\Phi_{10}(A_{11.5})^2$~erg~s$^{-1}$ where $\mu = \mu_{32}\times10^{32}$ G cm$^3$ is the magnetic dipole moment. 
The measured $\dot{P}$ gives $\mu\approx 1.6 \times 10^{32}$~G~cm$^3$, and one finds that the observed luminosity $L$ corresponds to $A\sim 3\times 10^{11}$~cm$^2$, which is close to the emission area of the thermal component of the \mgt\ spectrum. Thus, the thermal emission can be associated with the footpoint of the twisted bundle. It is bombarded by the relativistic particles from the $e^\pm$ discharge, and a fraction of power $L$ released in the bundle is radiated in a quasi-thermal
form at the footpoint, whose temperature may be estimated from
$A\sigma T^4\sim L$. This gives $kT \sim 1$~keV,
consistent with the observed thermal component. Additional heat
diffusion from the deeper crust could also contribute to the spot emission \citep{lyubar2002}, although
the available data do not yet require this.
%The complicated pulse profile suggests that the bundle is not axisymmetric
%and its hot footpoint may have a more complicated shape than a round polar cap.
The expected decay time of the magnetospheric luminosity is given by $t_{\rm ev}\sim 10^7 \mu_{32} \Phi_{10}^{-1} A_{11.5}$~s (Beloborodov 2009). \mgt\ is predicted in this model to show a decay time of $t_{\rm ev}\sim 10^{7}$~s, which is similar to that observed in XTE J1810-197. Additional monitoring with \swift\ can confirm our model prediction on the flux evolution \citep{kennea2013}.

%\subsection{\mgt\ in Sgr A* gravitational field}

In spite of the angular proximity of Sgr A* to the magnetar and the possibility that the latter is in the Galactic Center, we would not expect measurable bias in the spin-down rate due to orbital acceleration unless the orbit were very eccentric and the orbital phase highly fortuitous, and/or the currently measured spin-down rate were temporarily much larger than the intrinsic value.
Additionally, if at the Galactic Center, 
the magnetar is $\sim1.5^{\prime\prime}$ outside a disk with mean eccentricity $\sim0.3$ \citep{belobor2006,lu2009} of clockwise rotating, predominantly massive O-type and Wolf-Rayet (WR) stars with disk age $\sim6$ Myr \citep{geg2010}.
It would not be surprising that the magnetar would stay well localized if it were born in the disk. 
The disk escape velocity is $\sim750$ km~s$^{-1}$, much larger than measured magnetar velocities (e.g. \citet{tendulkar2012}).  
Moreover, depending on the precise details of initial eccentricity and orbital phase, a kick velocity $\sim100-200$ km~s$^{-1}$ could move the magnetar out of the disk and into an even more elliptical, bound orbit. 
We need not, however, specify the argument to the disk. 
Approximately 50\% of the O/WR stars, also of age $\sim6$ Myr, reside on extended, isotropic orbits outside the disk, and the magnetar would be similarly bound in that case. There are other stars in this region. B-stars have a density ~3-4 times lower than the early-type stars though. The density of much less massive late type stars is  $\sim$2 times higher. 
Thus plausibly associating the magnetar with the early-type stars, as has been previously suggested \citep{ritche2010}, would imply a progenitor mass $\goe 40M_{\odot}$, based on the age of all the early-type stars in this region, and evolutionary models.

\section{Conclusion} 

We report the \nustar\ discovery of a new magnetar known as \mgt. 
The detection of spin period and its first derivative confirms that \mgt\ is a transient magnetar previously undetected. \mgt\ is the fourth of a growing subclass of magnetars detected in the radio band. It indicates that many magnetars in quiescence remain undetected in the X-ray band or they have been detected as high-B radio pulsars. 
Further monitoring of \mgt\ by X-ray and radio telescopes will reveal the time evolution of the spectral and timing properties, thus constraining the outburst emission mechanism of transient magnetars. 
This discovery of a magnetar near Sgr A* has important implications for the dynamics, progenitor masses and formation of neutron stars in the vicinity of the Galactic Center, and these issues will be addressed in our follow-up paper.

%Associating the magnetar with a WR star of $\sim6$ Myr would imply a progenitor of mass  $\goe40M_{\odot}$, based on evolutionary models \citep{mam1987}.
%This is consistent with the few other estimates of magnetar progenitor mass. 
%(BUT MY QUESTIONS REMAIN -- ARE THERE NO LATER TYPE STARS IN THE VICINITY THAT ARE HARDER TO OBSERVE??)

\acknowledgements

This work was supported under NASA Contract No. NNG08FD60C, and made use of data from the \nustar\ mission, a project led by the California Institute of Technology, managed by the Jet Propulsion Laboratory, and funded by the National Aeronautics and Space Administration. The authors wish to thank Arash Bodaghee and Clio Sleator for their assistance with data analysis and Brian Metzger for helpful discussions.

%\bibliography{sgra_magnetar}

\end{document}